\documentstyle[12pt,twoside,fleqn,espcrc1,epsfig]{article}

\newcommand{\be}{\begin{eqnarray}}
\newcommand{\ee}{\end{eqnarray}}

\newcommand{\AmS}{{\protect\the\textfont2
  A\kern-.1667em\lower.5ex\hbox{M}\kern-.125emS}}

\title{QCD phases at high density and instantons}

\author{E.Shuryak\address{Department of Physics and Astronomy, State University of New York,      Stony Brook, NY 11794-3800}
}

\begin{document}
\maketitle

\begin{abstract}
The talk is an introduction into diquark condensation phenomena which
occur in QCD at high energy density. They are driven by instantons and
instanton-antiinstanton pairs (or ``molecules''), which generate attraction in
some qq channels. A number of phases is possible, with or without
restoration of chiral symmetry: the work is not finished and we do not yet
know which take place in real QCD. We also
emphasize that specific diquark correlations play a significant role
in baryon structure, in particular making that of a nucleon very
different of a $\Delta$ (or other member of a decuplet). This
``small $N_c$'' scenario based on comparison to QCD with two colors
is contrasted with the ``large  $N_c$'' one.

\end{abstract}

\section{INTRODUCTION}
At this workshop one probably does not need to explain why we study
QCD at large density: it is, after all, a traditional domain of
nuclear physics. However, we now  discuss it  in quite different
perspective, 
as a  part of recent development
related  with the QCD transitions
into what is generally known as the
quark-gluon plasma (QGP). 

  We know much more
about  high temperature QCD and the corresponding transitions, mostly 
 because of systematic use of our   main
tool, the
 lattice QCD. Numerical simulations have shown that 
 the main non-perturbative
features of the QCD vacuum - confinement and chiral symmetry breaking
- do indeed  disappear in a common 
 high temperature phase transition, at
 $T=T_c\sim 150 MeV$. The debates about its mechanism is a
 focal point of non-perturbative QCD: all vacuum models (using instantons,
 abelian
projected monopoles or  $Z_N$ vortices or any other beasts) claim to explain 
it. The same is claimed by many field theory models (e.g. the NJL
model) or even the resummed
perturbative one-gluon exchange!
However, in order to see which claims are in fact true
many significant details of the picture 
 should still be  clarified. Let me on the onset of the talk mention few
of them, first at qualitative and then quantitative level.

 Among qualitative questions people mostly discussed whether
their favorite objects mentioned above show any change at $T\approx T_c$.
I think the main issues to address now  are
 dependence of the finite T transitions
on the number of quark flavors $N_f$ and their masses.
The boundary of
the unknown here is not so  far out: we do not know what
happens for chiral phase transition for as small number of
 of flavors as $N_f\sim 5$, where the usual instanton-based
mechanism of chiral symmetry breaking seems to fail
and instanton-antiinstanton ``molecules'' were predicted to
dominate even the vacuum
state
  (see \cite{SS_98}).
 What happens between this number and $N_f=11 N_c/2$ 
(where asymptotic freedom ends)   is still a matter of speculations.
Amusingly,  we now know 
the phase diagram of super-symmetric QCD much
better, due to famous works of Seiberg and collaborators. If anything,
we have learned from them that fermions are important, and can lead to
many different phases.
 Another set
of qualitative questions deal with hadrons/correlation functions 
at $T>T_c$.
In particular, there are intriguing deviations of pion-sigma
``screening masses'' from  vector ones: can those be  reproduced
by all those models?

  One can certainly ask lots of questions at the quantitative level,
  but still let me point out only one important point which is rarely
discussed. Why is the phase transition in ``quenched QCD'' ($N_f=0$)
 $T_{quenched}\sim 260 MeV$  is so much higher than  $T_{N_f=2}\sim 150 MeV$? 
How exactly the hadronic spectrum in these case/or fermionic
determinant for particular configurations
leads to this smaller critical temperature?  

 Much less work has been done in the field of finite density QCD.
Since
the main perturbative phenomena (such as screening and plasma
oscillations)
in dense quark matter has been worked out (see e.g. \cite{Shu_80}),
the field was essentially dormant for 2 decades. 
The lattice approach to it was much discussed at this meeting, and it is
probably still true that so far this approach does not work for finite 
chemical potential. The issue is usually discussed
in connection with complex fermionic determinant,
 to which standard
 Monte-Carlo algorithms do not apply, and people try to avoid it.
However, other methods also start from the vacuum ensembles, trying to
deduce properties of high density matter from them. Very small overlap
between the two leads to  enormous numerical problems: what is needed is
some way to simulate the high density matter by itself. 

 Being short of such method, we use the well-tested
analytic methods such as mean field
 (MF) and Random Phase Approximation (RPA).  They have done a good job
for
 the usual super-conductivity, or pairing
correlations in nuclei. 
As the Coulomb interaction between quarks 
of {\em different colors} is attractive, it was realized early on 
that Cooper pairs can be formed and condense, making
the quark plasma  a super-conductor. The magnitude of the 
corresponding gap $\Delta$ and the critical temperature $T_c$ were 
estimated to be in the 1 MeV range~\cite{BL}.

 Recent interest to this problem was revived  by two groups
\cite{RSSV,ARW} which independently came from two different angles to a very similar
conclusions. They have shown that {\it instantons} may generate much stronger
``color super-conductor'', with a gap in the 100 MeV range. Specifically,
these works dealt with the simplest 2-flavor theory, in which only
one basic type of diquarks, I=0 or (ud-du) can be formed, in anti-triplet
color representation. After the original papers further progress include
  interesting new phases including strange quarks
 \cite{ARW2}. The issue here is whether a gap of similar
 magnitude can be generated
by instantons, or concentrate instead on  smaller  
gap generated by the one gluon exchange (OGE).

New rapidly growing field or research is related to
 the vicinity of the tricritical point \cite{BR,stephetal,SRS}. Those
 resemble the well known liquid-gas nuclear transition much studied
in low energy heavy ion collisions. Hopefully, it will create equally
fruitful experimental works!

  At the end of the introduction let me say
that we now  see only the tip of the iceberg.
 Diquark condensation is fighting against the chiral condensate
of different nature, and probably also against the mesonic condensates
discussed previously \cite{mesonic_cond}.  On top of it there appear 
instanton clusters (molecules), another kind of clustering which is
also
fermion-driven but do not need $any$ condensates at all. These three phenomena 
 make a triangle of love-hate
relationships not less complicated than in a classic novel. It will
take much more work to sort it out.

\section{WHY INSTANTONS?}
 
  Instantons are classical solutions of Yang-Mills equations, describing
the tunneling paths
from one minimum of the potential to the other.
As discussed e.g. in  recent review   \cite{SS_98},
instantons are the strongest non-perturbative phenomena in QCD.
In particularly, when lattice configurations are made somewhat
more smooth (removing quantum noise corresponding to perturbative 
high-frequency fields), it is instantons which they mainly see. 
Therefore, their density
is  
related to the value of the gluon condensate. Furthermore, it is
instantons
which seem to
generate the famous 1 GeV ``chiral Lagrangian'' scale\footnote{
 Note that this scale was for
a
long time  puzzling to people doing perturbative QCD: at 1 GeV nothing
bad happens with the perturbative running coupling constant.}
, separating perturbative
and non-perturbative domains (see recent
discussion of this issue in\cite{RRS}). Instanton-based models quantitatively
explain chiral symmetry breaking and pion properties, and even spectroscopy
of lowest mesons and baryons (including $N,\Delta$). Although they
by no means include all non-perturbative phenomena (e.g. they produce too
weak confining potential), they seem to be sufficient for description of all
phenomena relevant for traditional nuclear physics. 

 Instantons are present not only in QCD, and arguments for
their dynamical role are not only empirical.  As a good theoretical example
let me mention  N=2 SUSY QCD, for which  Witten and Seiberg 
have found $exact$ effective Lagrangian \cite{WS}. As the photon sector of
the theory has 
only one-loop perturbation theory, the Witten-Seiberg solution in weak coupling is
nothing but a series of the instanton-generated diagrams.
Instantons  blow up the coupling constant at a scale 
$2^{3/2}\Lambda_{pert}$, with an
 amusing numerical coincidence with real
QCD\cite{RRS} .

The main reason instantons are important in QCD is because of their
relation to anomaly and light fermions. A general topological argument
demands that tunneling produce certain rearrangement in fermionic sector:
this leads to effective interactions  we use.
It is very important that  these general arguments are
also valid in high density matter.
 In particular, the famous explanation of the anomaly as
fermion level motion (known also as an ``infinite hotel'' story)
can be directly applied to cold quark matter. An obvious conclusion
is that ``extra'' fermion pairs produced by tunneling must show up
in the same multiplicity as in vacuum, but now at
the Fermi surface\footnote{
 In random matrix models people have used a simple-minded approach: adding
 $\mu \gamma_4$ to the Dirac operator {represented as a random matrix with
some density of zero modes} (leading to quark condensate). If $\mu$ is
sufficiently large, eigenvalues move away from zero and chiral symmetry gets restored. 
}. This consideration alone suggests interesting
instanton-induced effects at high  density.
\section{QCD WITH ONLY TWO COLORS  }
  We start with this   theory for several reasons. 
The main one is pedagogical: it will teach us some important lessons about
diquark spectroscopy. The secondary one is that it is a very special
case, in which simple theoretical and even lattice studies of the
finite $\mu$ case can be done. 
This theory is in many ways 
  the opposite to the large $N_c$ limit, and thus is very
instructive to keep it in mind.  For large $N_c$ 
baryons are supposed to be $N_c$ times heavier than mesons. In the $N_c=2$ QCD 
baryons (or {\em diquarks})
 are in fact
{\em degenerate} with mesons,
 due to Pauli-Gursey symmetry\footnote{
Unlike in SUSY, in this case there are different number of baryons and
mesons
in multiplets, and both are of course bosons.} 
(PGSY) \cite{PG},
 relating quarks and anti-quarks.
 In particular, the lowest baryons (or diquark) should be  bound as
 strongly
as  the lowest meson, the massless pions. 
 The general pattern of symmetry breaking is 
$SU(2N_f)\rightarrow Sp(2N_f)$  and the number of Goldstone modes is  
$N_{goldstones}=2 N_f^2 - N_f -1$.
Let us mention two cases specifically.
For $N_f=1$ there remains $no$ goldstones because
there is no symmetry to be spontaneously broken. (Recall that due to U(1)
  anomaly, even the corresponding meson, $\eta'$, is massive.)
 For the most
interesting case  $N_f=2$ the coset (ratio) of full group over
remaining one is 
\be K=SU(4)/Sp(4)=SO(6)/SO(5) =S^5\ee
which means the 5-dim sphere with 5 massless modes: 
three of
those are pions, plus scalar diquark S and its anti-particle $\bar S$.
  At finite but small density one finds that
 rotations on the 5-dim sphere  cost no
energy, and  by doing so 
 in the direction of the scalar diquark condensate one naturally obtains
states with non-zero diquark density. So the critical point is $\mu=0$
(for a non-zero quark mass it is half the pion mass), and the usual chiral
condensate $<\bar q q>$ is rotated into a diquark one. 
 The qualitative picture can be understood 
using the corresponding linear sigma model. The potential
\begin{equation} 
 V = \lambda \left( \vec\pi^2 \!+\! \sigma^2 \!+\! S^2 \!+\! \bar S^2
     \!-\! v^2 \right)^2 - A\sigma - \mu^2 \left( S^2 \!+\! \bar S^2 \right)
\end{equation} 
includes the diquark chemical potential $\mu$ and the chirally asymmetric 
mass term $A$. At $\mu=0$ the Goldstone masses are $m_g^2=A/v$, and 
$m_\sigma^2=8\lambda v^2$. For non-zero $\mu$ we can determine the 
$\langle\bar qq\rangle$ and $\langle qq\rangle$ condensates 
$\langle\sigma\rangle$ and $\langle S\rangle$ using the mean 
field approximation. We find
\begin{equation} 
 4\lambda \langle S \rangle \left( \sigma^2+\langle S \rangle^2-v^2\right)
 =2\mu^2 \langle S \rangle \ .  
\end{equation} 
Below the critical chemical potential $\mu_c\simeq m_g/\sqrt{2}$, 
$\langle\sigma\rangle$ is constant and $\langle S\rangle=0$. Above
$\mu_c$, $\langle S\rangle$ increases as 
\begin{equation} 
 \langle S \rangle^2 = {\mu^2\over 2\lambda} + v^2 - {A^2 \over 4 \mu^2} 
\end{equation}
and $\langle \sigma \rangle = m^2_g v/(2\mu^2)$. The energy density is 
$\epsilon =-\mu^2v^2-{3 A^2/(4\mu)^2}$, compared to $\epsilon =-m_g^2
v^2+m_g^4/(16 \lambda)$ for the normal vacuum. 

  Unlike real QCD, $N_c=2$ gauge theory is straightforward to simulate 
on the lattice, since the fermion determinant remains real for $\mu\neq 0$. 
With the exception of some early work using small lattices and the strong
coupling expansion \cite{strongcoupling}, few studies have taken 
advantage of this. Numerical studies of the instanton model for $N_c=2$
at finite density have  been done  in\cite{S_97}.
All of them display how
 at large density the chiral $\langle\bar qq\rangle$ condensate is 
replaced by the diquark one $\langle qq\rangle$,
in agreement  with
 the sigma model described above. 
\section{DIQUARKS IN REAL QCD (WITH 3 COLORS)}
Before we return to the real world, we review properties of the instanton-induced
interaction due to 't Hooft, rewritten   
 in $\bar q q$ and $q q $ channels.
We consider only the simplest case of
two flavors
(up and down). As a shorthand notations, effective
 Lagrangian can be written as follows
\be 
\label{l_mes}
{\cal L} &=& G{1\over 8 N_c^2}
 \left[(\bar\psi \tau^- \psi)^2+
 (\bar\psi \tau^- \gamma_5 \psi)^2 \right],
\ee
where we have added the interaction in the direct and exchange 
channels and dropped color octet terms. $N_c$ is the number of 
colors and $\tau^-=(\vec\tau,i)$ is an isospin matrix. In this way we
have combined isospin 1 channel (including the pion) with the isospin 0 one
(denoted as $\eta'$). The i squared leads to
a sign difference for them, showing that the same interaction
tends to make pion light and $\eta'$ heavy.
The coupling G is related to the instanton density and we would not discuss
it here, except to comment that
the huge magnitude of the $\pi-\eta'$ splitting
 hints once again about large scale of the
instanton-induced effects.

 In this discussion we are actually
interested in another manifestation of this Lagrangian,
in the diquark channel. Its
phenomenological implications  in 
this case were first discussed in connection with spin-dependent 
forces in baryons \cite{BL_SR}, challenging the conventional wisdom that 
spin splittings in baryons are due to the one-gluon exchange\footnote{
Of course, even its strong proponents should explain where magnetic
moment
of a quark comes from. Obviously it must be the same mechanism which
generates the constituent quark mass, and therefore we are back to instantons.
}. 
  The same Lagrangian (\ref{l_mes}) can be Fierz-rearranged into a $(qq)$ 
interaction:
\be 
\label{l_diq}
{\cal L} &=& 
G \left\{
 -{1\over 16 N_c (N_c-1)}
 \left[ (\psi^T C \tau_2 \lambda_A^n \psi)
        (\bar\psi\tau_2 \lambda_A^n C \bar\psi^T) 
  \right.\right.\\
  & &  \left. \hspace{2.0cm}\mbox{}
       +(\psi^T C \tau_2 \lambda_A^n \gamma_5 \psi)
        (\bar\psi \tau_2 \lambda_A^n \gamma_5 C \bar\psi^T) \right]  
               \nonumber\\
  & &  \left. \mbox{}
       +{1\over 32 N_c (N_c+1)}
        (\psi^T C \tau_2 \lambda_S^n \sigma_{\mu \nu} \psi)
        (\bar\psi \tau_2 \lambda_S^n \sigma_{\mu \nu} C \bar\psi^T) 
        \right\} \nonumber
\ee 
Here, $C$ is the charge conjugation matrix, $\tau_2$ is the anti-symmetric 
Pauli matrix, $\lambda_{A,S}$ are the anti-symmetric (color $\bar 3$) and 
symmetric (color 6) color generators. This Lagrangian (\ref{l_diq}) 
provides a strong attractive interaction between an up and a down quark with 
anti-parallel spins ($J^{P}=0^{+}$) in the color anti-triplet channel, and
a repulsive interaction in the $0^-$ channel. $0^+$ quark pairs couple to 
the diquark current $S^a_{dq} = \epsilon_{abc} u^T_b C \gamma_5 d_c$.

 Quantitative studies of instanton effects 
in baryon spectroscopy, both for light and heavy-light systems, were done 
in~\cite{SSV}.
 It was first found that
 instanton interaction alone not only makes the nucleons and deltas to be
 bound states of constituent quarks, but even
their splitting  also came out right (being even somewhat larger
than the  observed one). Further studies have found that  the nucleon 
 has a very large overlap with 
the current $\epsilon_{abc}(u^T_aC\gamma_5d^b)u^c=S^a_{dq}u^a$,
containing scalar diquark. 
Both observation were clarified by further analysis, in which 
either the third quark is taken to be infinitely heavy (and thus
contributing
only the color phase matrix), or fixing the gauge and ignoring the
third quark completely. Either way one finds a deeply bound
scalar diquark (Sdq) $2m_q-m_{Sdq}\simeq 200-300$ MeV, 
 whereas in
all other channels (vectors and axial-vectors, color 6 diquarks, etc.)
no such binding was found. 
Furthermore, one can derive the diquark binding analytically,
in the usual RPA approximation\footnote{The opposite
conclusion reached in \cite{DFL}  is due to summation of
only part of the needed diagrams.}.  The first lattice attempt
\cite{Karsch}  to study ``diquark spectroscopy'' have indeed
found that a scalar diquark is more  
 bound than vector one, but the effect is rather modest\footnote{We
 hope future data at smaller quark masses will provide more accurate
extrapolation to physical masses: the
 present data (plus linear extrapolation in quark mass) 
gives only about half of the nucleon-delta mass splitting. Note also,
that the qualitative difference between nucleon and delta $correlators$ at
intermediate distances found in the instanton model
\cite{SSV} were confirmed on the
lattice by the MIT group \cite{CGHN_93b}.}.

 Note  a nice continuity at this point:
while changing from $N_c=2$ to 3 colors the
same scalar diquark 
goes from being nearly massless (bound by about $2 M_{const}\sim 700 MeV$)
to a deeply bound state. It can  be traced to
 the coefficients of the
 Lagrangian given above:  for $N_c=2$ the
corresponding coupling constants in mesonic and diquark channels are
equal,
while they become factor 2 different  for $N_c=3$.

 Let us qualitatively compare the ``large
$N_c$'' picture of baryons with our ``small $N_c$''.
 The former
suggests a picture of baryons as large heavy objects, made of
$classical$
pion fields. In it $N,\Delta$ (and other members of octet and
decuplet) are basically the same object, slowly rotating with a slightly
different angular momenta. 
The latter picture is radically different: the octet baryons can be
pictured
as a
strongly bound (and spatially compact) scalar diquark plus the third quark.
The decuplet baryons do not have scalar diquarks,
 and are therefore generically
 3-body objects. 
Which one is closer to the real world? 
A long list of phenomenological hints suggesting the quark-diquark
picture is better can be found in a review \cite{diquarks}.

  One particularly important observation is strikingly different behavior
 of the electromagnetic form-factor of the
 $N\rightarrow \Delta$ transition, as compared to
$N\rightarrow N^*$ (and also elastic one): The former is decreasing with
$Q^2$ much quicker than the latter. It is difficult to get such behavior,
unless the structure of $N$ and $\Delta$ are very different.

Let me add one more hint to the list\footnote{I am indebted to M.Strikman,
who brought this interesting works to my attention.}.
 Although in order to be convinced by
it one  still  has some experimental work to do,
 it is rather general and model-independent.
  It is related to a phenomenon of {\it inelastic diffraction} in high
energy hadronic collisions. As noticed by Good and Walker already in
1960, it exists due to different absorption probability of different
component in hadronic wave functions. (If all of them be absorbed equally, 
the only diffraction left would be the elastic one.) Data on $\pi$
and N inelastic diffraction of nuclear targets have been analyzed
 in \cite{BFS}, and the corresponding   distributions $P(\sigma)$  over 
fluctuating cross sections $\sigma$ were derived. Amazingly, the one
for the nucleon is as wide (if not wider) than that for the pion. 
 Let me speculate that this wide distribution
 is due to quasi-two-body (quark-diquark) structure of a
nucleon. The  
 opposite large-$N_c$-inspired picture,  a skyrmion, should on the contrary
lead to small fluctuations of essentially classical object\footnote{ Large observed
fluctuations of $\sigma$ also contradict to another a priori possible  picture,
is that of a large
number N
of partons. If those were to
interact independently, one should  expect relatively
small fluctuations of $\sigma$, $\sim 1/N^{1/2}$.}.
Can one check this interpretation experimentally? Unfortunately there
can be no beams of $\Delta$ particles. However, there are those
for another decuplet member, the celebrated $\Omega^-$. What size are
fluctuations of $its$ cross section on nuclei? Is it really
smaller
than that for the nucleon? 

\section{INSTANTON EFFECTS AT HIGH DENSITY: VARIOUS PHASES}

  The problem we are going to discuss is an ensemble
of quarks coupled to an ensemble of instantons. 
Like in so many other
problems of statistical mechanics, we find one more case of a competition
between order and disorder, or energy versus entropy.  
The QCD vacuum has strongly disordered instantons, with nearly random
distributions and non-zero chiral condensate. 
Mean field analysis seem to be quite adequate in this case:
it suggests a simple picture that at whatever space-time point the coupling
occurs, the fermion ``legs" of the Lagrangian given above can be simply
absorbed by the vacuum condensates.
Curiously
enough, the QGP phase
 at
 high T is made of $ordered$ (or clustered) instantons,  combined into 
so called instanton-anti-instanton molecules \cite{IS,SSV}. 
They have been predicted to be directed in time direction, and there
are first lattice data \cite{DeForcrand} showing that at
$T\sim T_c$ this is indeed the case.
There are no condensates
available in the usual QGP phase, and
 so (massless) quarks has to propagate from
instanton to anti-instantons. 

\begin{figure}[h]
\epsfxsize=3.0in
\centerline{\epsffile{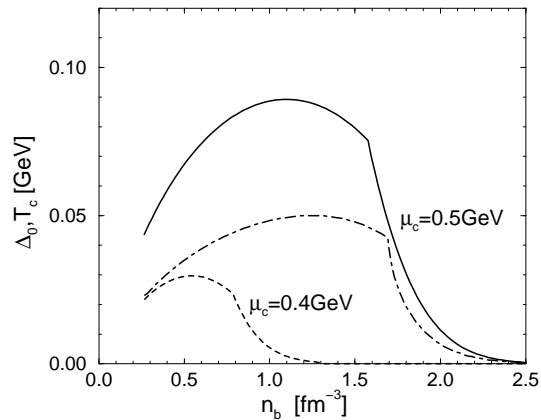}}
\caption{The gap $\Delta(\mu,T$=0) for $\mu_c$=0.4,~0.5~GeV (dashed and full
line, respectively) and critical temperature $T_c$ (dash\-dotted line,
$\mu_c$=0.5~GeV) versus baryon charge density $n_b$.  }
 \label{fig_gap}
\end{figure}
  The situation is different at high density, because new type of condensate,
the diquark ones, can be formed. For example, with two flavors, all 4 legs of
the instanton vertex can  be absorbed by diquark and anti-diquark from a
condensate. This means that instanton ensemble may remain to be``random",
with the mean field approximation applicable. This is what was done by  
\cite{RSSV,ARW} who get a very similar
conclusions. The gap and critical temperature are shown in Fig. 1.
This condensate is colored, and thus it breaks color SU(3) to SU(2):
however it preserves chiral symmetry. In the mean field approximation
one possible phase, that of chirally asymmetric gas of constituent quarks,
is unstable: so quark matter with diquark condensate appears in a first
order transition directly from the vacuum.
 Compared to the original papers mentioned,
 further work is being made
to incorporate the realistic instanton form-factors
(see talk by Velkovsky) and include
instanton-anti-instanton molecules (see talk by Rapp).

  The situation becomes more interesting if one includes 
the strange quark\footnote{
Since the critical $\mu_c\sim 300-350 MeV$ is larger than $m_s\approx
140 MeV$,
strange quarks
 should be included in realistic calculations relevant for neutron
 stars. }.
The instanton-induced interaction becomes a sum of (i) the $4-fermion$
$(\bar u u) (\bar d d)$ interaction with the coefficient proportional to
$m_s$ and (ii) the $6-fermion$ interaction, with  
all 3 flavors. If one ignores the former one (or puts $m_s=0$), there is no way
to put all 6 legs into 
 diquark condensates. So we are forced to invent something new.

  Alford, Rajagopal and Wilczek (ARW) \cite{ARW2} have decided to abandon
instantons  and return to
the one-gluon exchange (OGE) mechanism\footnote{  The results of ARW are 
numerically  large, suggesting a gap comparable to that induced by
instantons.  It would be in general surprising to have such
strong effect of a
perturbative  nature, and I hope it will be  reduced after further
scrutiny. It seems that there was a wrong numerical factor in overall 
 normalization of the coupling constant, which enter in the exponent.
}. They have found that the energy minimum corresponds
to the phase with
 ``diagonal" locking of color (Latin indices) and flavor (Greek ones)  
\be
< q^T_{i\alpha}  C \gamma_5 q_{j\beta}> =K_1 \delta_{i\alpha}\delta_{j\beta}+K_2 \delta_{i\beta}\delta_{j\alpha}  
\ee
This phase preserves some diagonal combination of color and flavor,
but (in spite of absent $<\bar q q>$)
it is  {\it chirally asymmetric}. As a result there are massless pions, which
can (as usual) be obtained by acting by the chiral generators on the condensate
given above. These pions look like diquark rather than quark-anti-quark states,
but baryon number is broken anyway.

 Let us however return to the instanton-induced 6-fermion interaction. The only
ways to get a non-zero answer in the mean field approximation is to put
its 6 legs either into the
combination of the condensates $<\bar q  q>^3$
(like in vacuum) or into   $<qq><\bar q \bar q><\bar q  q>$.
The latter demands $both$ types of condensates to be present, so we are
now forced to break the chiral symmetry.
 It follows however  from the structure of
the corresponding gap equations that this phase cannot be present in a weak
coupling
regime: like the NJL model, this solution exists only above certain
 critical coupling. 

  This unwanted feature however disappears when
we go outside the usual mean field approximation and include
the instanton-anti-instanton
``molecules''. Let me provide some motivation on why should we do it.
 In order to explain why these configurations are  important, 
let us compare their contribution with
  one would typically obtain instead in the mean field (MF) approximation. 
In both cases we discuss
 the contribution to the partition function is a second order
diagram in the basic 't Hooft interaction, with $(2N_f)$ fermion lines
going between two vertices. In the MF treatment
the correlation between color orientations of $I$ and $\bar I$ is
ignored,
so that an  averaged coupling is used. Crude estimate of the
inaccuracy of this step can be made, even ignoring the orientation
dependence of the gauge action (the dipole forces of
Callan-Dashen-Gross). Looking at fermionic ``hopping''
amplitudes one finds that each of them contains explicit factor
cos($\theta$), where $\theta$ is the so called relative orientation angle. 
The amplitude for ``molecules'' contains the factor
$<(cos(\theta))^{2N_f}>$,
  and therefore the integral is dominated by the region close to  $\theta=0$, the more so the larger
is $N_f$. So, the angular integral is much larger than the power of its average
value $(1/4)^{N_f}$  used in the MF.

 Clustering of instantons ($I$) and anti-instantons ($\bar I$) into pairs
 is the simplest
possible ``ordering''.
If it happens, quark wave functions get localized (inside these pairs) and
  chiral symmetry gets restored. Both analytic \cite{IS,SV} and
numerical simulations of the instanton ensemble \cite{SS_95} show
that at high T and/or large $N_f$ the ensemble indeed
breaks into such objects.
  In the diquark problem at high density these ``molecular''
  configurations are especially important,  because for $any$ value of $N_f$
they can provide 
 the desired {\it 4-fermion}
effective interaction: all ``unwanted''
$(2N_f-4)$ lines can in this case be internal ones. If the relative
orientation is locked at  $\theta=0$ (as explained above), it is
described by a simple universal Lagrangian derived and discussed in \cite{SSV}. 

The coupling constant (proportional to a density of such molecules)
can only be evaluated by some explicit integration and is model
dependent. However the structure of the Lagrangian is unique.
 It is instructive to compare it with the original 't Hooft
  interaction (for 2 flavors) and OGE. Like two others, it may only
  contain
chirally symmetric operators, and (unlike 't Hooft) it should also respect
chiral U(1) symmetry because the ``molecules'' are topologically
trivial. For mesonic channels the effective Lagrangian reads
\be 
L_{molecular}^{mesonic}= {2 G\over N_c^2} [ (\bar q \tau^a q)^2 -  (\bar q \gamma_5\tau^a q)^2\\
+ (\bar q \gamma_\mu\gamma_5 q)^2 -(\bar q \tau^a\gamma_\mu\gamma_5
q)^2/4-(\bar q \tau^a\gamma_\mu q)^2/4] + (color_octets)] \nonumber
\ee
while for color-anti-triplet diquarks it is 
\be   
L_{molecular}^{ \underline 3 diquarks} = {G\over N_c(N_c-1)} [ 
|(q^T C \gamma_5 t^A \lambda_A q)|^2 - |(q^T C  t^A \lambda_Aq)|^2 \\
+ |(q^T C \gamma_\mu\gamma_5 t^A \lambda_A q)|^2/4 - 
|(q^T C \gamma_\mu t^A \lambda_A q)|^2
] \nonumber
\ee
Only  anti-symmetric flavor and color generators  $\lambda_A,t_A$ are
present here (the
symmetric part is not shown, it has no scalar diquarks we are
interested in.) For $N_c=3$ the diquarks are factor 2 weaker, as in the
 t'Hooft
Lagrangian.
For comparison, OGE leads to mesonic and diquark Lagrangian
to the coefficients $(K/4)(N_c^2-1)/N_c^2$ in the pion-sigma channels,
and $(K/8)(N_c+1)/N_c$ for flavor-color-antisymmetric diquarks and  
 $-(K/8)(N_c-1)/N_c$ for flavor-color-symmetric diquarks. 
The detailed calculations are not yet finished, but the preliminary result
is that a phase with diagonal color-flavor locking (similar to that obtained
in \cite{ARW2} using a OGE interaction) happens to be the lowest one.
  
 At very high density 
 instantons are Debye screened by quarks 
\cite{Shu_82}. Therefore, one should expect in this limit to
return to the OGE interaction and the gaps and the critical
temperature
in the MeV range. 
So if the above-mention conclusion
would survive, the whole high density domain would be in the
``diagonal'' phase.

  Whether the simpler (and chirally symmetric) ``non-diagonal'' phase
discussed above for $N_f=2$ theory does or does not take place at some
intermediate densities  depends on the value of the strange
quark mass $m_s$. If it is big enough, one would have the 2-flavor
chirally symmetric phase, with only ud condensate being large. If it is
smaller than critical value,  then we are in the diagonal 3-flavor phase
in which  
chiral symmetry  remains broken at any density, without change of phase.

\section{ONE MORE POSSIBLE PHASE, IN WHICH DIQUARKS ARE $NOT$ COOPER PAIRS} 
 Let us return to the low density limit of real QCD with three colors, 
ignoring strangeness. (For definiteness, we consider neutron ($udd$) matter 
relevant for stars, in which $ud$ diquarks and $d$ 
quarks compensate each others color and electric charges.)
Let us ask the following question. If one accepts the existence of deeply
bound diquarks inside nucleons, they have the smallest energy per
unit baryonic charge, and one may ask whether the nuclear matter (made of
nucleons)  wins energetically\footnote{In QCD at low density the answer
would be positive simply due to confinement, which forced diquarks to
be connected by color strings. However we study the instanton model,
 in which the only interaction between quarks is mostly the
 quasi-local
't Hooft one, and gauge-induced potential is relatively weak. }
 over a gas of such
diquarks.

  The first observation is that Bose condensation of diquarks at zero T
is inevitable, and
since scalar diquarks are color anti-triplets, the condensate will
select a direction in color space. If we label this direction red ($k$=3),
our $ud$ diquarks are made of blue and green ($k$=1,2) quarks only.
The third quark-type (red $d$) is basically unaffected, and thus it should
form a Fermi gas.
So even in the most naive scenario (all diquarks condense in zero momentum
state so that their energy is just the sum of masses) one has to pay a price
in form of Fermi motion of the 3-rd quark. It is easy to see that it is in fact
larger than Fermi motion of nucleons, and so the nuclear matter is indeed the 
lowest low-density state.

Can a diquark gas still be better at some intermediate density? 	It is
possible to make such naive models in which it is the case.
But even with strong binding and including gains due to diquarks  
 Bose character, one cannot definitely answer this question.

\begin{figure}[h]
\epsfxsize=4.0in
\centerline{\epsffile{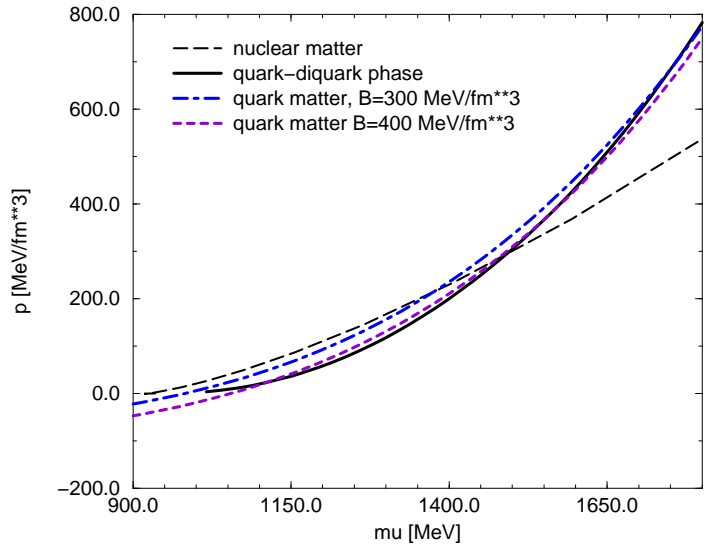}}
\caption{ Pressure versus baryonic chemical potential
(3 times $\mu$ for quarks used above) for 3 phases: the {\em nuclear
  matter} made of nucleons, the {\em quark-diquark} phase made of Bose gas of
  interacting
diquarks and Fermi gas of constituent quarks (no bag
constant), and {\em Fermi gas} of u,d,s
quarks with current masses (with two different bag constants indicated).  }
\label{fig1_fb}
\end{figure}

Clearly it would be erroneous 
to conclude that an infinite number of diquarks condenses in the $p=0$ 
state: diquarks are composite objects and, like nucleons, they
should have a short-range repulsion.
 One may account for repulsion by introducing a 
scattering length (below $a\simeq 0.3$ fm). Using well known expression for 
 the interacting Bose gas, one can get its energy (per quark)  
\begin{equation}
{\epsilon^B \over n_q}={4\pi a n_S \over m} 
\left( 1+ {128 \over  15\pi^{1/2}}(a^3 n_S)^{1/2}\right)
\end{equation}
where $n_S$ is the scalar diquark density. The first term is the mean 
field result, and the second term comes from non-condensed diquarks
\cite{LY}. Results of our calculations are 
shown in  Fig.\ref{fig1_fb}, in which ``quark-diquark matter''
 consists of the following components: (i) a Bose 
gas of $S$ diquarks in chemical equilibrium with (ii) a Fermi gas of 
blue/green quarks, and neutralized in color and electric charge
by an appropriate amount of (iii) red quarks. For definiteness, we 
use the diquark and quark masses of 500~MeV and  400~MeV, respectively.

\section{SUMMARY}
We have argued that instanton-induced interaction leads to strong
color superconductivity in dense matter, with a gap of the order of
100 MeV. We think it is unlikely that heavy ion collisions can reach
the corresponding part of the phase diagram, and so the main
application should be dense stars. 

Two different symmetries of the condensates compete: a ``non-diagonal''
solution,
in which some direction in color space is spontaneously selected, or 
``diagonal'' one, in which color is locked with flavor. The former
solution preserves chiral symmetry, the second breaks it. It is not
yet clear whether we have both of them subsequently, as
density grows,
 or only the second one. The answer depends on numerics, in
 particularly on the relation between the empirical value of the strange quark
 mass and its critical value, separating those phases.

One more phase is possible (although this predictions is not robust),
 containing diquark Bose gas. If it exists, it should separate quark matter
 from nuclear one. Unfortunately, both this phase and nuclear
 matter
itself can only be discussed with more refined theoretical methods
than the mean field analysis applied so far.

\section{Acknowledgements} This talk is based  
 on a contuing work done in collaboration with R.~Rapp,
 T.~Sch{\"a}fer and M.~Velkovsky. Partial support by the US DOE
                (grant DE-FG02-88ER40388) is also acknowledged. 

{\small

\end{document}